\documentclass[prd,nofootinbib,twocolumn]{revtex4-1}

\usepackage{graphicx,epsfig}
\usepackage{amsmath}
\usepackage{amssymb}
\usepackage{dsfont}
\usepackage{pifont}

\def\be{\begin{equation}}
\def\ee{\end{equation}}
\def\beq{\begin{eqnarray}}
\def\eeq{\end{eqnarray}}

\def\A{{\cal{A}}}
\def\dal{\text{\ding{113}}}

\begin{document}

\title{On Derrick's theorem in curved spacetime.}

\author{Sante Carloni}
\email{sante.carloni@gmail.com}
\affiliation{Centro de Astrof\'isica e Gravita\c c\~ao - CENTRA,
Departamento de F\'isica,
Instituto Superior T\'{e}cnico - IST,
Universidade de Lisboa - UL,
Avenida Rovisco Pais 1, 1049-001, Portugal}

\author{Jo\~{a}o Lu\'{i}s Rosa}
\email{joaoluis92@gmail.com}
\affiliation{Centro de Astrof\'isica e Gravita\c c\~ao - CENTRA,
Departamento de F\'isica,
Instituto Superior T\'{e}cnico - IST,
Universidade de Lisboa - UL,
Avenida Rovisco Pais 1, 1049-001, Portugal}

\date{\today}

\begin{abstract} 
We extend Derrick's theorem to the case of a generic irrotational curved spacetime adopting a strategy similar to the original proof. We show that a static relativistic star made of real scalar fields is never possible regardless of the geometrical properties of the (static) spacetimes. The generalised theorem offers a tool that can be used to check the stability of localised solutions of a number of types of scalar fields models as well as of compact objects of theories of gravity with a non-minimally coupled scalar degree of freedom. 
\end{abstract}

\maketitle

\section{Introduction}\label{Introduction}
Derrick's theorem \cite{Derrick} constitutes one of the most important results on localised solutions of the Klein-Gordon in Minkowski spacetime. The theorem was developed originally as an attempt to build a model for non point-like elementary particles \cite{Wheeler:1955zz,Enz1963} based on the now well known concept of ``quasi-particle''. Wheeler was the first to suggest the idea of electromagnetic quasiparticle which he called {\it Geons}. In spite of the fact that Wheeler's geons do not really exist, other models were proposed (and are still studied) in which Geons are composed of other fields in various setting. There are even (time dependent) formulations of this idea which are based on gravitational waves \cite{Brill:1964zz}.

It is clear that in the exploration of the idea that fundamental particles could be some form of Geons, a crucial problem is to infer the stability of the Geon itself. Derrick's theorem deals specifically with the stability of Geons made of scalar fields. In particular, Derrick found that in flat spacetimes the Klein Gordon equation cannot have static solution with finite energy \cite{Manton:2004tk}.

In relativistic astrophysics, Derrick's theorem has profound consequences: its proof implies that no stable boson star can be constructed with real scalar fields and therefore that the existence of these objects requires more complex fields. Indeed the term boson stars nowadays is largely used to refer to complex scalar field stars, which are also called $Q-balls$ \cite{Coleman}. 

The consequences of Derrick's result span many different field of physics from low energy phenomena to QCD, to non linear phenomena, to pure mathematics (see e.g. the list of papers citing \cite{Derrick}). This is due to the fact that Derricks's results is related to a very general property of a class of differential equations called ``Euclidean scalar field equations'' to which the static Klein-Gordon equation belongs. In particular Derrick's theorem is a direct consequence of the so called Pohozaev identity \cite{Pohozaev,Berestycki}. This identity is akin to the well known Virial theorem as it relates the kinetic and potential energy of a localised scalar field configuration.

The original Derrick's theorem is limited to the case of flat spacetime and since its publication a number of works have been published considering particular cases, metrics or matter fields (see e.g. \cite{Palmer,Radmore:1978ux,Heusler,Herdeiro:2015waa,Hod:2018dij,Herdeiro:2018wub,Alestas:2019wtw}). However, no general proof of this theorem in curved spacetime and backreaction has been given. The purpose of this work is to provide such generalisation. The proof is based on the use of the 1+1+2 covariant approach  \cite{Clarkson:2002jz,Betschart:2004uu,Clarkson:2007yp}. With this tool we will be able to extend Derrick's results to the case of a curved spacetime. We will also discuss the consequences of such results on compact objects in some types of modifications of General Relativity.

The paper is organised as follows: in Section II we will describe briefly the 1+1+2 formalism which will be the main tool of our proof. In Section III we will use the 1+1+2 formalism to prove Derrick's theorem in  flat spacetime. In Section IV we will extend this theorem to curved spacetimes. In Section V we will analyse the effect of backreaction on the results of Section IV. Section VI explores the effect that scalar field coupling might have on the proof of the extended Derrick theorem. Section VII is dedicated to the application of the generalised Pohozaev identity to different relevant models of scalar fields. Section VIII concerns the application of Derrick's theorem to non minimally coupled theories of gravity. Finally Section IX is dedicated to the conclusions.     

Unless otherwise specified, natural units ($\hbar=c=k_{B}=8\pi G=1$) will be used throughout this paper and Latin indices run from 0 to 3. The symbol $\nabla$ represents the usual covariant derivative and a comma corresponds to partial differentiation. We use the $-,+,+,+$ signature and the Riemann tensor is defined by
\begin{equation}
R^{a}{}_{bcd}=\Gamma^a{}_{bd,c}-\Gamma^a{}_{bc,d}+ \Gamma^e{}_{bd}\Gamma^a{}_{ce}-\Gamma^e{}_{bc}\Gamma^a{}_{de}\;,
\end{equation}
where the $\Gamma^a{}_{bd}$ are the Christoffel symbols (i.e. symmetric in the lower indices), defined by
\begin{equation}
\Gamma^a_{bd}=\frac{1}{2}g^{ae}
\left(g_{be,d}+g_{ed,b}-g_{bd,e}\right)\;.
\end{equation}
and $g_{ab}$ is the metric tensor. The Ricci tensor is obtained by contracting the {\em first} and the {\em third} indices
\begin{equation}\label{Ricci}
R_{ab}=g^{cd}R_{acbd}\;.
\end{equation}
Finally round brackets around indexes of a given tensors represent symmetrisation of these indexes, whereas square brackets represent antisymmetisation:
\begin{equation}\label{Symm}
\begin{split}
X_{(ab)}&=\frac{1}{2}\left(X_{ab}+X_{ba}\right)\;,\\
X_{[ab]}&=\frac{1}{2}\left(X_{ab}-X_{ba}\right)\;.
\end{split}
\end{equation}
                
\section{Some elements of the 1+1+2 covariant formalism}\label{sec112}
In what follows we will make use of the 1+1+2 covariant formalism \cite{Clarkson:2002jz,Betschart:2004uu,Clarkson:2007yp} to construct a proof of Derrick's theorem in curved spacetime and in the context of modified gravity. 

In the 1+1+2 formalism a generic spacetime is foliated in 2 surfaces, which we will call $\Upsilon$, by the definition of a timelike and a spacelike congruence represented by the vector $u_a$ and $e_a$ respectively. The metric tensor can then be decomposed as 
\begin{equation}
g_{ab}=-u_au_b+e_ae_b+N_{ab},
\end{equation}
where $N_{ab}$ is, at the same time a projector operator and the metric of $\Upsilon$. It will be useful also to define a 3 surface $W$ with metric $h_{ab}=e_ae_b+N_{ab}$.

In line with the above decomposition we can define three differential operators: 
a dot $\left(\ \dot{}\ \right)$ represents the projection of the covariant derivative along $u_a$ e.g.
\be
\dot{X}^{a..b}{}_{c..d}{} = u^{e} \nabla_{e} {X}^{a..b}{}_{c..d}~,
\ee
a hat $\left(\ \hat{}\ \right) $ denotes the projection of the covariant derivative along $e_a$ e.g. 
\be
\hat{X}_{a..b}{}^{c..d} \equiv  e^{f}D_{f}X_{a..b}{}^{c..d}~,
\ee
$ \delta_a$ represents the covariant derivative projected with $N_{ab}$ e.g. 
\be
\delta_lX_{a..b}{}^{c..d} \equiv  N_{a}{}^{f}...N_{b}{}^gN_{h}{}^{c}...
N_{i}{}^{d}N_l{}^jD_jX_{f..g}{}^{h..i}\;.
\ee
At this point the kinematics and dynamics of any spacetime can be described via the definition of some specific quantities constructed with the derivatives of $u_a$, $e_a$ and $N_{ab}$. 

If one consider a spacetime endowed with a local rotational symmetry (LRS) i.e. a spacetime in which a multiply-transitive isometry group acting on the spacetime manifold, the 1+1+2 formalism allows to write the equations in terms only of scalar quantities. In our case only the quantities
\be\label{112Quant}
\begin{split}
&\mathcal{A}=e_a u_b\nabla^b{u}^{a}=e_a \dot{u}^{a}~,  \\
&\phi = N_{ab}\nabla^b e^{a}= \delta_a e^a~,\\
&\mathcal A_b=N_{ab}\dot{u}^{a},\\
& a_b= \hat{e}_b,\\
& \zeta_{ab}=\left( N^{c}{}_{(a}N_{b)}{}^{d} - \frac{1}{2}N_{ab} N^{cd}\right) \nabla_{c}e_{d},
\end{split}
\ee
will be necessary. 

Notice that our treatment will not involve vorticity as vortical spacetimes are inherently stationary and we are interested here only in static spacetimes. In the following, for sake of simplicity, we will call such general irrotational spacetimes ``curved''. 

It is important to clarify the limits of the approach that we will follow to  extend Derrick's idea. We first assume that our curved spacetime is such that at any point the quantities $u_a$, $e_a$ and $N_{ab}$ can be defined as $C(1)$ tensor fields. In other words the spacetime must be regular enough to be consistent with those fields. 

\section{Covariant Derrick theorem in flat spacetime}\label{secDTF}
The equation of motion for a real scalar field $\varphi$ minimally coupled to gravity is a Klein-Gordon equation of the form
\be\label{KleinGordon}
\dal\varphi-V_{,\varphi}=0,
\ee
where $\dal=\nabla^a\nabla_a$ is the D'Alembert operator, $\nabla_a$ is the covariant derivative with respect to the metric $g_{ab}$, $V=V\left(\varphi\right)$ is the scalar field potential, and $V_{\varphi}$ denotes a derivative with respect to the scalar field $\varphi$.

In \cite{Derrick}, to explore stability one expresses the variation of the action deforming the spatial coordinates with a constant parameter $\lambda$ in the Klein Gordon action. We will use here the properties of the covariant approach to perform an equivalent operation. For simplicity, let us consider first the spherically symmetric case in a Minkowski spacetime. In the covariant language, a deformation like the one used by Derrick in \cite{Derrick} can be represented by the {\it quasi-conformal} transformation
\begin{equation}\label{QCBeta}
\begin{split}
& u_a\Rightarrow \bar{u}_a=u_a,\\
& e_a\Rightarrow \bar{e}_a=\frac{1}{\lambda} e_a,\\
& N_{ab} \Rightarrow \bar{N}_{ab}=\frac{1}{\lambda^2} N_{ab},
\end{split}
\end{equation}
where $\lambda$ is assumed to be a generic positive function. 
Under \eqref{QCBeta} the D'Alembertian of $\varphi$
\begin{equation}\label{dA_Def}
\dal\varphi= \frac{1}{\sqrt{-g}}\partial_a\left(\sqrt{-g}g^{ab} \varphi_{,b}\right)_{,a}\;,
\end{equation}
transforms as
\be\label{dA_Transf}
\dal\varphi\Rightarrow \lambda^2\dal\varphi-\lambda \lambda^{,a}\varphi_{,a}\,,
\ee
which in static flat spacetime can be written as
\begin{equation}\label{DeformBeta}
\begin{split}
&\dal\varphi\Rightarrow \lambda^2\varphi_{,qq}-\lambda \lambda_{,q}\varphi_{,q}\,,
\end{split}
\end{equation}
where $q$ is a parameter associated to the congruence $e_a$. Using the relation above, Eq. \eqref{KleinGordon} becomes
\be\label{Klein112FlatSSSDef}
\lambda^2\varphi_{,qq}-\lambda \lambda_{,q}\varphi_{,q}-V_\varphi=0.
\ee
Equations of this type do not satisfy in general the Helmholtz conditions \cite{Starlet} and therefore they cannot be directly obtained as the Euler-Lagrange equations of any Lagrangian. However, Darboux showed \cite{Darboux,Baldiotti} that in one dimension there is an equivalent second-order equation for which a variational principle can be found, namely,
\be\label{Klein_Helmotz}
e^{\Phi}\left(\lambda^2\varphi_{,qq}-\lambda \lambda_{,q}\varphi_{,q}-V_\varphi\right)=
0,
\ee
where $e^{\Phi}$ is known as the {\it integrator multiplier}. The form of the integrator multiplier in the case of an equation with the structure of \eqref{Klein_Helmotz} can be found via the relation
\be\label{EqPhi_SS}
\frac{d}{d\varphi_{,q}}\mathcal{Q}- \frac{d}{d q}\left[\frac{d}{d\varphi_{,qq}}\mathcal{Q}\right]=0\,,
\ee
where $\mathcal{Q}$ represents \eqref{Klein_Helmotz} and we assumed that $\Phi$ does not depend on the derivatives of $\varphi$. In our case it turns out that $\Phi=-3\ln \lambda+ \Phi_0$ where $\Phi_0$ is a constant. We will choose here $\Phi_0=0$ so that $\Phi=0$ for $\lambda=1$ and we recover the original action. With this choice the action for \eqref{Klein112FlatSSSDef} is given by  
\be\label{Action112FlatSSD}
\begin{split}
S(\lambda)=&-\frac{1}{2}\int \frac{1}{\lambda^3}\left[\lambda^2\varphi_{,q}^2+2V\left(\varphi\right)\right]dq\,.
\end{split}
\ee
If the solution of \eqref{EqPhi_SS} is localised, this integral will be well defined and finite. Derrick's deformation is given by 
\begin{equation}
\lambda= const.
\end{equation}
which implies that \eqref{Action112FlatSSD} can be written as  
\begin{align}\label{Action112FlatDD}
 S(\lambda)=
&-\frac{1}{2}\left(\frac{I_1}{ \lambda}+\frac{I_2}{\lambda^3}\right),
\end{align}
where 
\begin{equation}\label{I1_I2_defFlatDD}
\begin{split}
I_1&=\int \varphi_{,q}^2\, d q \;,\\ 
I_2&=2\int V\left[\varphi(q)\right]\, d q\;.
\end{split}
\end{equation}
Following \cite{Derrick}, we now check that Eq.~\eqref{KleinGordon} corresponds to an extremum of the action requiring that  
\be
\frac{\partial S(\lambda)}{\partial \lambda}=0 \rightarrow \frac{I_1}{ \lambda^2}+3\frac{I_2}{\lambda^4}=0.
\ee
Setting $\lambda=1$, we obtain that 
\be \label{DActFlat}
I_2=-\frac{I_1}{3},
\ee
i.e. the Pohozaev identity. This relation tells us that the Klein Gordon equation can be an extremum of action \eqref{Action112FlatDD} only if the integral of the potential is negative. This implies that, for example, a mass potential, which is defined positive, would never lead to an equilibrium.

We can determine the character of the extremum by considering the second order derivative of $S(\lambda)$:
\be
\frac{\partial^2 S(\lambda)}{\partial \lambda^2}=-\frac{I_1}{ \lambda^3}-6\frac{I_2}{\lambda^5}.
\ee
Substituting the \eqref{DActFlat} and setting $\lambda=1$ we obtain
\be
\frac{\partial^2 S(\lambda)}{\partial \lambda^2}=I_1>0.
\ee
Hence Eq.\eqref{KleinGordon} is a minimum for the action provided that the integral of the potential $V$ is negative.   

Now, in the static case the energy function (as defined in \cite{GoldsteinBook}) of $\varphi$ can be related to the action via the relation\footnote{This relation can be easily verified calculating directly the $(0,0)$ of the stress energy density for the scalar field which corresponds to the Hamiltonian, or more precisely, to the Lagrangian energy.} 
\be\label{EproptoS}
E= -2S,
\ee
 which implies
\be
\frac{\partial^2 E(\lambda)}{\partial \lambda^2}\bigg|_{\lambda=1}=-2I_1<0\;.
\ee
Therefore a minimum of the action corresponds to a maximum of the energy, and a localised  solution $\varphi(q)$ of the Eq.\eqref{KleinGordon} must be unstable. 

We can generalise this reasoning to the non spherically symmetric case in which also $\delta$ derivatives appear. From the \eqref{dA_Def}, using the parameters $w_2$ and $w_3$ to map the 2-surface $\Upsilon$, the d'Alembertian can be written as
\begin{equation}
\begin{split}
 \dal\varphi\Rightarrow &\lambda^2\varphi_{,qq}-\lambda \lambda_{,q}\varphi_{,q}+\\
 &+\sum_{i=2}^{3}\left(\lambda^2\varphi_{,w_i w_i}-\lambda \lambda_{,w_i}\varphi_{,w_i}\right).
 \end{split}
\end{equation}
Hence, the Klein-Gordon equation is
\be\label{Klein112FlatS}
\begin{split}
\lambda^2\varphi_{,qq}-&\lambda \lambda_{,q}\varphi_{,q}+\\ 
&\sum_{i=2}^{3}\left(\lambda^2\varphi_{,w_i w_i}-\lambda \lambda_{,w_i}\varphi_{,w_i}\right)-V_\varphi=0.
\end{split}
\ee 
Considering the  equivalent equation   
\be\label{Klein112FlatSDef}
\begin{split}
e^{\Phi}\Bigg[ & \lambda^2\varphi_{,qq}-\lambda \lambda_{,q}\varphi_{,q}\\ &~~~+\sum_{i=2}^{3}\left(\lambda^2\varphi_{,w_i w_i}-\lambda \lambda_{,w_i}\varphi_{,w_i}\right)-V_\varphi\Bigg]=0,
\end{split}
\ee
the integrator multiplier can be calculated using a condition similar to\footnote{Here we appear to force the original approach by Darboux, which works only for one dimensional actions. However we will show in Appendix~\ref{app} and in the following sections that the integrator multiplier can be associated to the volume form for the scalar field action and therefore it can be determined with the Darboux procedure also in multidimensional actions (at least in our specific case). Indeed it will become clear that the form of the integrator multiplier is actually irrelevant for our purpose, because its transformation properties can be deduced in general.} \eqref{EqPhi_SS}:
\begin{equation}\label{EqPhi_NSS}
\sum_{i=1}^{3}\left\{\frac{d}{d\varphi_{,p_i}}\mathcal{Q}- \frac{d}{d p_i}\left[\frac{d}{d\varphi_{,p_i  p_i}}\mathcal{Q}\right]\right\}=0\;,
\end{equation}
where $p_i=(0,q,w_2,w_3)$, $\mathcal{Q}$ is \eqref{Klein112FlatSDef} and we assumed again that $\Phi$ does not depend on the derivatives of $\varphi$. This relation amounts to the partial differential equation
\begin{equation}
\sum_{i=1}^{3}\left[\Phi_{,p_i}+3\frac{\lambda_{,p_i}}{\lambda}\right]=0.
\end{equation}
Using the method of the characteristics we can find the solutions 
\begin{equation}
\Phi=- 3\ln\lambda + C(w_2-q,w_3-q) \;.
\end{equation}
Since we want  to return to the standard action $e^\Phi=1$ for $\lambda=1$ we can set $C=0$.
Thus the action can be written as
\be\label{Action112FlatD}
S=-\frac{1}{2}\int \frac{1}{\lambda^3}\left[\lambda^2\varphi_{,q}^2+\lambda^2\sum_{i=2}^{3}\varphi_{,w_i}^2+2V\left(\varphi\right)\right]d \Omega\,,
\ee
where the $d\Omega=\prod_{a=1}^{3} d p_i$. Repeating the procedure above, we obtain
\begin{align}\label{Action112FlatDDNSS}
 S(\lambda)=
&-\frac{1}{2}\left(\frac{I_1}{ \lambda}+\frac{I_2}{\lambda^3}\right),
\end{align}
where 
\begin{equation}\label{I1_I2_defFlatDDNSS}
\begin{split}
I_1&=\int \left[\varphi_{,q}^2+\sum_{i=2}^{3}\varphi_{,w_i}^2\right]\, d\Omega, \\ 
I_2&=2\int V\left[\varphi(q)\right]\,  d\Omega,
\end{split}
\end{equation}
which implies
\be \label{DActFlatNSS}
\begin{split}
&\frac{\partial E(\lambda)}{\partial \lambda}\bigg|_{\lambda=1}=0\rightarrow I_1=- 3I_2, \\
&\frac{\partial^2 E(\lambda)}{\partial \lambda^2}\bigg|_{\lambda=1}=-2I_1<0,
\end{split}
\ee
and shows that the solution $\varphi$ is unstable. 
 
This result is called in literature Derrick's theorem and it is the main reason why localised solutions of real scalar fields are generally  considered unphysical.  In the next sections we will give a generalisation of this result in the case of irrotational LRS spacetimes and explore its validity in more general spacetimes and in the context of modified gravity.

\section{Covariant Derrick theorem in curved spacetimes}\label{secIII}
Let us now prove Derricks theorem in curved spacetimes. As before, for simplicity we will start with the spherically symmetric case and then we will consider more complex cases.
\subsection{Spherically symmetric spacetimes}

Decomposing Eq.\eqref{KleinGordon} in the 1+1+2 variables and considering spherically symmetric LRSII spacetimes, the transformed Klein-Gordon equation reads
\be\label{Klein112CurvSSDef}
\lambda^2\varphi_{,qq}-\lambda \lambda_{,q}\varphi_{,q}+\left[\mathcal A(\lambda)+\phi(\lambda)\right]\lambda\varphi_{,q}-V_\varphi=0.
\ee
The above equation can be generated by the action
\be\label{Action112CurvSS}
S(\lambda)=-\frac{1}{2}\int e^{\Phi(\lambda)}\left[\lambda^2\varphi_{,q}^2+2V\left(\varphi\right)\right]dq\,,
\ee
where 
\be\label{Phi_Beta}
\Phi(\lambda)=\int\left\{\frac{1}{\lambda}\left[\mathcal A(\lambda)+\phi(\lambda)\right]-3\frac{\lambda_{,q}}{\lambda}\right\}dq \,.
\ee
The integral \eqref{Phi_Beta} can be simplified remembering that the under transformation \eqref{DeformBeta} 
we have
\be\label{TransAPhi}
\mathcal{A}(\lambda)=\lambda e_a\dot{u}^a=\lambda \mathcal{A}\,,\qquad
\phi(\lambda) =\lambda \delta_a e^a =\lambda\phi \,.
\ee 
This means
\be\label{SolPhiSS}
\Phi(\lambda)-\Phi_0=\int\left[\mathcal A+\phi\right]dq- 3\ln \lambda \,,
\ee
which yields
\be \label{ExpPhiTransfSS}
e^{\Phi(\lambda)}\Rightarrow\frac{e^{\Phi}}{\lambda^3}\,,
\ee
where we have chosen $\Phi_0=0$ so that $\lambda=1$ implies $e^{\Phi(\lambda)}=e^{\Phi}$. 
In this way equation \eqref{Klein112CurvSSDef} can be derived from the action
\be\label{Action112CurvSSFin}
S=-\frac{1}{2}\int \frac{e^{\Phi}}{\lambda^3}\left[\lambda^2\varphi_{,q}^2+2V\left(\varphi\right)\right]dq\,.
\ee
The above expression is consistent with the interpretation of $e^{\Phi}$ as the volume form for the action \eqref{Action112CurvSS} of equation \eqref{Klein112CurvSSDef} (see Appendix~\ref{app} for details). In this perspective the choice that we made for $\Phi_0$ corresponds to a choice of the asymptotic properties of the metric. This fact can be understood bearing in mind that, by definition, $\A$  and $\phi$ are identically zero when the spacetime is Minkowskian \cite{Carloni:2013ite}. As we consider localised solutions for $\varphi$, it is only natural to choose an ``asymptotically flat'' $\Phi$ by choosing $\Phi_0=0$.

Setting $\lambda=const.$, we can write
\begin{equation}\label{Action112CurvedDD}
 S(\lambda)=
-\frac{1}{2}\left(\frac{I_1}{ \lambda}+\frac{I_2}{\lambda^3}\right),
\end{equation}
where this time 
\begin{equation}\label{I1_I2_defCurvSSFin}
\begin{split}
I_1&=\int e^{\Phi}\varphi_{,q}^2\, d q, \\ 
I_2&=2\int e^{\Phi}V\left[\varphi(q)\right]\, d q.
\end{split}
\end{equation}
Action \eqref{Action112CurvedDD} is the same of \eqref{Action112FlatSSD} and leads to the same conditions. This implies that Derrick's theorem is valid also in the curved spherically symmetric case. 

It is important to stress that the Darboux procedure we have used so far to deduce the action is valid only if the integrator multiplier is different form zero. One can prove \cite{Carloni:2013ite} that this condition implies that the spacetime we are considering does not contain a perfect or Killing horizon. Such constraint excludes the case of spacetimes describing black holes, trapped surfaces etc.
\subsection{General irrotational spacetimes}

What about more complex spacetimes? If vorticity is zero, 
upon the transformations \eqref{DeformBeta}  the Klein-Gordon equation  reads
\be\label{Klein112CurvedS}
\begin{split}
\lambda^2&\varphi_{,qq}-\lambda \lambda_{,q}\varphi_{,q}+\left[\mathcal A(\lambda)+\phi(\lambda)\right]\lambda\varphi_{,q}+\\ 
&+\sum_{b=2}^{3}\left[\lambda^2\varphi_{,w_i w_i}+\lambda \lambda_{,w_i}\varphi_{,w_i}+\mathcal A_b(\lambda)\lambda^2\varphi_{,w_i}\right.
\\
&\left.+a_b(\lambda)\lambda^2\varphi_{,w_i}\right]-V_\varphi=0.
\end{split}
\ee
It is clear that in non spherical irrotational LRS spacetimes where $\mathcal A_b$ and $a_b$ are identically zero (which still belong to the LRSII class), Derrick's theorem holds. We have
\begin{align}\label{Action112CurvedDDNSS}
 S(\lambda)=
&-\frac{1}{2}\left(\frac{I_1}{ \lambda}+\frac{I_2}{\lambda^3}\right),
\end{align}
where 
\begin{equation}\label{I1_I2_defCurvedDDNSS}
\begin{split}
I_1&=\int e^{\Phi}\left[\varphi_{,q}^2+\sum_{i=2}^{3}\varphi_{,w_i}^2\right]\, d\Omega\,, \\ 
I_2&=2\int e^{\Phi}V\left[\varphi(q)\right]\, d\Omega\,,
\end{split}
\end{equation} 
 and $\Phi$ is given by the \eqref{Phi_Beta} for $\lambda=const.$

If we consider more general spacetimes ($\mathcal A_b\neq0$ and $a_b\neq0$), we have to explore the transformation of the acceleration vectors under \eqref{QCBeta}. We have
\begin{equation}\label{TransAba}
\begin{split}
 \mathcal A_b(\lambda)=&N_{b}{}^c \dot{u}_c=\mathcal A_b,\\
  a_b(\lambda)=&N_b{}^c \hat{e}_c=N_b{}^c \lambda \widehat{\left(\frac{e_c}{\lambda}\right)}=a_b.
\end{split}
\end{equation}
Defining the four vector
\be
V_a =(\A+\phi)e_a+\left(\A_c+a_c\right)N^c_b,
\ee
condition \eqref{EqPhi_NSS} for this case takes the form
of the partial differential equation involving the components of $V_a$
\begin{equation}\label{EqDarboux112Curv}
\sum_{i=1}^{3}\left[\Phi_{,p_i}+3\frac{\lambda_{,p_i}}{\lambda}+ V_i\right]=0\,.
\end{equation}

We can use the method of characteristics to solve the above equation and, as in all partial differential equations, the existence and properties of the solutions will depend critically from the boundary conditions. As we have seen in the spherically symmetric case (see also Appendix \ref{app}) the boundary conditions are strictly related to the  asymptotic properties of the specific metric which one is considering. Since we have assumed that the scalar field is localised, it is natural to assume asymptotic flatness. However, as far as it generates the correct field equation the exact form of $\Phi$ is irrelevant for our purposes. We only need to determine the transformation of the quantity \eqref{ExpPhiTransfGEN} under \eqref{QCBeta}. From \eqref{EqDarboux112Curv} it is evident that $\Phi$ will transform such that 
\be \label{ExpPhiTransfGEN}
\Phi(\lambda)=\Phi -3\ln\lambda.
\ee  
Using the above result, we obtain the action 
\begin{align}
 S(\lambda)=
&-\frac{1}{2}\left(\frac{I_1}{ \lambda}+\frac{I_2}{\lambda^3}\right),
\end{align}
where 
\begin{equation}
\begin{split}
I_1&=\int e^{\Phi}\left[\varphi_{,q}^2+\sum_{i=2}^{3}\varphi_{,w_i}^2\right]\,  d\Omega\,, \\ 
I_2&=2\int e^{\Phi}V\left[\varphi(q)\right]\,  d\Omega\,.
\end{split}
\end{equation} 
This is the same result obtained in the flat case.

\section{Introducing backreaction}
In the previous section we have made the tacit assumption that the mass of the confined scalar field solution would not perturb the assigned metric of the spacetime. In other words we have neglected {\it backreaction}. 

Is it possible to generalise the strategy above to the case in which the localised scalar field solution also contribute to the spacetime metric?
In this case, one should add the Hilbert Einstein term to the action for the scalar field
\begin{equation}\label{eq:actionMCPhi}
\begin{split}
S=\frac{1}{2}\int d x^{4}\sqrt{-g}\Big[R&
-\nabla_a\varphi\nabla^a\varphi 
- 2V(\varphi)
\Big]\;,
\end{split}
\end{equation}
and derive its transformation under \eqref{QCBeta}.

It should be pointed out that at present there is no general consensus on the definition of the energy of the gravitational field. One should ask, then, if it makes sense to extend Derrick's results also to the backreaction case. A positive answer can be provided thinking that we are considering a very special case.  First of all, in order to keep finite the integral \eqref{Action112CurvedB}, we have to assume an asymptotically flat background. In addition, since our choice of the vector field $u_a$ corresponds to a timelike Killing field for the spacetimes we consider, the class of observers we consider is static. 

From the results of \cite{Katz:2006uw}  we have that in stationary spacetimes the energy of the gravitation field can be written as the scalar
\begin{equation}\label{EqGGen}
E_G=\int \sqrt{-g} \left(t^m{}_n u^n+ \sigma^{[mn]}{}_p\partial_n u^p\right)u_m d\Omega, 
\end{equation}
where $t^m{}_n $ is the Einstein pseudotensor and $\sigma^{[mn]}{}_p$ is Freud's complex \cite{Freud} given by
\begin{equation}
\sigma^{[mn]}{}_p= \frac{1}{g}g_{pr}\left(g\, g^{r [m}\,g^{n] s}\right)_{,s}.
\end{equation}
The \eqref{EqGGen} can be written, in our assumptions, as
\begin{equation}
E_G= -\frac{1}{2}
\int \sqrt{-g} \mathcal{L}^{(3)}_{\bar{\Gamma}\bar{\Gamma}} d\Omega,
\end{equation}
where
\begin{equation}
\mathcal{L}^{(3)}_{\bar{\Gamma}\bar{\Gamma}}=h^{ab}\left(\bar{\Gamma}_{a d}{}^{c}\bar{\Gamma}_{c b}{}^{d}-\bar{\Gamma}_{ab}{}^{c}\bar{\Gamma}_{c d}{}^{d}\right),
\end{equation}
the $\bar{\Gamma}$ being the Christoffel symbols of the three surface $W$.  Now, starting from the Hilbert-Einstein action in the static case we can write
\begin{equation}
S_G= \frac{1}{2}\int \sqrt{-g}\,R\, dt d\Omega=\frac{T_0}{2}\int \sqrt{-g}\,R \,d\Omega,
\end{equation}
where $T_0$ is a constant, which we can set to one without loss of generality. Using the contracted Gauss-Codazzi equation, we have
\begin{equation}\label{SpiltR4}
\begin{split}
S_G= \frac{1}{2}\int \sqrt{-g}\Big(&R^{(3)}+ K^2 - K_{a}{}^{b}K^{a}{}_{b}+\\
&\nabla_a\left[\dot{u}^a+ u^a K\right]\Big) d\Omega.
\end{split}
\end{equation}
where $R^{(3)}$ is the Ricci scalar for  submanifold $W$,
\be
K_{ab} = h_a{}^c h_b{}^d \nabla_{c}u_{d},
\ee
is the second fundamental form of $W$ and $K=K^{a}{}_{a}$. Using the Gauss theorem we can integrate out the last factor in \eqref{SpiltR4}. In addition, in static irrotational spacetimes the terms associated to the extrinsic curvature are identically zero. Now, $R^{(3)}$ can be decomposed in as similar way as $R$ in \eqref{SpiltR4}, employing the extrinsic curvature of $\Upsilon$. Integrating out the second (projected) derivatives and rewriting the expression in terms of the Christoffel symbols, we arrive to 
\begin{equation}\label{E_G}
S_G=  S_{0}+\frac{1}{2}
\int \sqrt{-g}\,\mathcal{L}^{(3)}_{\bar{\Gamma}\bar{\Gamma}}\,d\Omega= -\frac{E_0}{2}- E_G.
\end{equation}
where $S_0$ is a constant and we have defined $E_0=-2S_0$. Thus in our case (and only in this case), modulus an irrelevant constant, the energy of the gravitational field  can linked to the Hilbert Einstein action:
\begin{equation}
E_G= -\frac{E_0}{2}-\frac{1}{2}\int e^{\Phi}R \;d\Omega.
\end{equation}
Here, using the derivation of Appendix~\ref{app} we have connected the volume form to $e^\Phi$.

The next task is to evaluate how $E_G$ transforms under \eqref{QCBeta}. Using the Gauss Codazzi equation also on $R^{(3)}$ gives, in static irrotational spacetimes, 
\be
R =R^{(2)}-2 \hat{\phi}-\frac{3}{2}\phi^2- 2 \zeta ^2 -2 a_b a^b+ 2\delta_b a^b.
\ee
Now, since  
\be
R^{(2)}=2K_G,
\ee
where $K_G$ is the gaussian curvature, the Brioschi formula implies that, under \eqref{QCBeta},
\be
 R^{(2)}\Rightarrow R^{(2)}(\lambda)=\lambda^2 R^{(2)}+...
\ee
where the dots represent terms which contain derivatives of $\lambda$. In addition, from definition \eqref{112Quant} one finds
\be
\begin{split}
\zeta_{ab}&\Rightarrow\zeta_{ab}(\lambda)=\frac{\zeta_{ab}}{\lambda }\\
\zeta&\Rightarrow\zeta(\lambda)=\lambda^2 \zeta.
\end{split}
\ee 
Using also the \eqref{TransAPhi} and \eqref{TransAba} we arrive at
\be\label{R_transf}
 R (\lambda)=\lambda^2 R +...
\ee
where, again, the dots represent additional terms which contain derivatives of $\lambda$. As we will eventually set up $\lambda=1$, these terms are irrelevant and can be neglected.

The total action derived from \eqref{eq:actionMCPhi} transforms as 
\be
\begin{split}\label{Action112CurvedB}
&S(\lambda)=\frac{S_G}{\lambda}+\\&-\frac{1}{2}\int \frac{e^{\Phi}}{\lambda^3}\Bigg[ \lambda^2\varphi_{,q}^2
+\lambda^2\sum_{i=2}^{3}\varphi_{,w_i}^2+2V\left(\varphi\right)\Bigg] d\Omega\,.
\end{split}
\ee
In the case $\lambda=const.$, defining
\begin{equation}\label{I1_I2_I3defB}
\begin{split}
I_1&=\int e^{\Phi}\left[\varphi_{,q}^2+\sum_{i=2}^{3}\varphi_{,w_i}^2\right]\,  d\Omega , \\ 
I_2&=2\int e^{\Phi}V\left(\varphi\right)\,  d\Omega,\\
I_3&=-E_0-2 E_G, 
\end{split}
\end{equation}
one can write
\begin{equation}\label{Action112CurvedBLam}
E(\lambda)=-2 S(\lambda)=
\left(\frac{I_1}{ \lambda}-\frac{I_3}{ \lambda}+\frac{I_2}{\lambda^3}\right),
\end{equation}
Since, as we have seen, relation \eqref{EproptoS} between the energy and the action is still valid, we can examine the stability of the backreacting solution with the same strategy as the previous section. We have  
\be
\begin{split}\label{Stab_Back_GR}
&\frac{\partial E(\lambda)}{\partial \lambda}\bigg|_{\lambda=1}=0\rightarrow I_3=I_1+ 3I_2,\\
&\frac{\partial^2 E(\lambda)}{\partial \lambda^2}\bigg|_{\lambda=1}=6I_2\,.
\end{split}
\ee
Now the trace of the field equations provides another relations that should be taken in consideration. We have 
\be\label{TraceEFE_GR}
R=\nabla_a\varphi\nabla^a\varphi+4V,
\ee
i.e., upon integration,
\be\label{TraceEFE_GR_I}
I_3=I_1+2I_2.
\ee 
Combining the above results with the first of \eqref{Stab_Back_GR} gives $I_2=0$. This yields 
\be\label{Derrick_B}
\frac{\partial^2 E(\lambda)}{\partial \lambda^2}\bigg|_{\lambda=1}=8I_2=0.
\ee
Since the above quantity has opposite signs if we consider $I_2\rightarrow I_2\pm\epsilon$, where $\epsilon$ is a small constant, we have an inflection. Hence the presence of gravity has weakened the instability but cannot eliminate it completely. 

The weakest point of the reasoning given above is, undoubtedly, the definition of the gravitational energy of the system. One might object that even with our specific assumptions the definition of energy we have used might miss some crucial aspect of the physics of these systems. We can argue here that this is not the case going around the problem of the definition of $E_G$ by eliminating the Hilbert Einstein term from the action using the field equations i.e. considering the {\it on shell} action. 
  
For example, using the relation \eqref{TraceEFE_GR} we have   
\be\label{Action112CurvedB_2}
E^{tot}(\lambda)=-2\int \frac{e^{\Phi}}{\lambda^3}V\left(\varphi\right) d\Omega\,,
\ee
which immediately implies the result \eqref{Derrick_B}. This result shows that our previous argument is correct and at the same time suggests a easy shortcut to prove Derrick's theorem with backreaction. In the following we will make ample use of this shortcut, especially in dealing with more complex settings. 

We can use the on shell action to probe further in the validity of Derrick's theorem, by considering, for example, the case in which the scalar field backreacts with a spacetime with non zero cosmological constant $\Lambda$. The \eqref{Action112CurvedBLam} now reads
\begin{equation}\label{Action112CurvedBLam_CC}
E(\lambda)=
\left(\frac{I_1}{ \lambda}-\frac{I_3}{ \lambda}+\frac{I_2}{\lambda^3}+\frac{2 I_4}{\lambda^3}\right),
\end{equation}
where
\be
I_4=\Lambda \int e^{\Phi}  d\Omega.
\ee
We obtain, on shell,
\be
\begin{split}\label{Stab_Back_GR_CC}
&\frac{\partial E(\lambda)}{\partial \lambda}\bigg|_{\lambda=1}=0\rightarrow I_3=I_1- I_2,\\
&\frac{\partial^2 E(\lambda)}{\partial \lambda^2}\bigg|_{\lambda=1}=3(I_1+2I_2)\,.
\end{split}
\ee
Hence in this case stability is possible if
\be
I_2>-\frac{1}{2}I_1.
\ee
Therefore the presence of a cosmological constant can lead to stable solutions. However these solutions make sense physically only at scales in which $\Lambda$ is relevant, and therefore exclude microscopic or astrophysical objects. Yet, the picture that emerges is that Derrick's instability cannot be avoided by minimal modifications of the model. In the following we will explore further the validity of Derrick's theorem looking at the effect of scalar field coupling, non canonical scalar field and modified gravity.

\section{The role of scalar field couplings}\label{secV}
Derrick's instability is very robust. No additional standard coupling of the scalar field with matter or other fields can prevent its appearance. A  coupling with another scalar field of the type $f(\varphi)g(\psi)$  would just make more complicated the definition of the integral $I_2$. In fact, starting from the corresponding Klein-Gordon equations 
\be\label{KleinCF}
\dal\varphi-V_{,\varphi}-f_{,\varphi}g(\psi)=0,
\ee
and proceeding as in the previous section we obtain 
\begin{align}\label{Action112CurvedCoupled}
 S(\lambda)=
&-\frac{1}{2}\left(\frac{I_1}{ \lambda}+\frac{I_2}{\lambda^3}\right),
\end{align}
where 
\begin{equation}\label{I1_I2_defCurvedCoupled}
\begin{split}
I_1&=\int e^{\Phi}\left[\varphi_{,q}^2+\sum_{i=2}^{3}\varphi_{,w_i}^2\right]\,  d\Omega\,, \\ 
I_2&=2\int e^{\Phi}\left\{V\left[\varphi(q)\right]+ f\left[\varphi(q)\right] g\left[\psi(q)\right]\right\}\,  d\Omega\,.
\end{split}
\end{equation} 
It follows that we can prove Derrick's theorem  also in this case. This conclusion is independent from the sign of the terms appearing in the above integral. The presence of the coupling, however, changes the physical significance of the Pohozaev equilibrium condition. The same happens when we introduce backreaction.

What about other types of coupling? The strategy of the proof we have presented shows that, whatever the coupling, the key point in the determination of the stability of localised scalar field configurations depends on the $\lambda$-dependence of the transformation of the integrator multiplier. If the transformation of $\exp(\Phi)$ is such that the action can be written as a combination of $\lambda$ terms and $\lambda$-independent integrals, like e.g. in \eqref{ExpPhiTransfSS}, there will be a chance to prove (in)stability. In other cases, Derrick's approach does not lead to a definite answer. 

A simple example is the case of  derivative coupling of the type $ a \hat{\varphi} g(\psi)$. For this coupling the transformation of the integrator multiplier is given by 
\be\label{Phi_g}
\begin{split}
 \Phi_{g}=\Phi+ a \int\varphi_{,q} \frac{g(\psi)}{\lambda^2} dq\,,
\end{split}
\ee
and $\lambda$ is not factorisable. This fact makes it impossible to find a form of the action similar to \eqref{Action112CurvedDDNSS}. 

Instead, considering a coupling of the type $ a \hat{\varphi} \hat{\psi}^2$ will yield
\be\label{Phi_gd}
\begin{split}
& e^{\Phi_{gd}(\lambda)}\Rightarrow\frac{e^{\Phi_{gd}}}{\lambda},\qquad \Phi_{gd}=\Phi+ a \int\varphi_{,q} (\psi_{,q})^2 dq\,,
\end{split}
\ee
which leads to an action similar to \eqref{Action112CurvedCoupled},  and thus implies instability.

\section{Non canonical scalar fields}
In the context of cosmology and in particular when dealing with the problem of dark energy a number of non canonical scalar fields have been introduced. Using the strategy above, we can extend Derrick's theorem to these cases. In the following, we will consider the case of: Phantom fields \cite{Phantom}, Quintom fields \cite{Quintom} and k-essence \cite{kessence}. In these models, as in the ones of next Section, for the sake of brevity, we will make Derrick's deformation directly in the action, rather than prove that the transformed action comes from the modified Klein-Gordon equation. This connection is, however, always valid. We will also consider only the case of LRSII spacetimes as the generalisation to more complicated geometries can be derived easily from the considerations above.

Phantom fields are scalar fields whose action contains a kinetic term with opposite sign with respect to the canonical one: 
\begin{equation}\label{eq:actionPh}
\begin{split}
S=\frac{1}{2}\int d x^{4}\sqrt{-g}\Big[R&
+\nabla_a\varphi\nabla^a\varphi 
- 2V(\varphi)
\Big]\;.
\end{split}
\end{equation}
Excluding backreaction, we have
\begin{align}\label{Action112CurvedPh}
 E(\lambda)=
&-\left(\frac{I_1}{ \lambda}-\frac{I_2}{\lambda^3}\right),
\end{align}
where
\begin{equation}\label{I1_I2_defCurvedPh}
\begin{split}
I_1&=\int e^{\Phi}\left[\varphi_{,q}^2+\sum_{i=2}^{3}\varphi_{,w_i}^2\right]\,  d\Omega\,, \\ 
I_2&=2\int e^{\Phi}V\left[\varphi(q)\right]\,  d\Omega\,.
\end{split}
\end{equation} 
Equation \eqref{Action112CurvedPh} yields
\be
\begin{split}\label{Stab_Ph}
&\frac{\partial E(\lambda)}{\partial \lambda}\bigg|_{\lambda=1}=0\rightarrow I_1= 3I_2,\\
&\frac{\partial^2 E(\lambda)}{\partial \lambda^2}\bigg|_{\lambda=1}=2I_1>0,
\end{split}
\ee
which implies that a localised solution of phantom fields is actually stable.  This result reveals that a key element of Derrick's instability is the sign of the scalar field  kinetic terms.

The inclusion of backreaction, however, introduces instability. On shell, the energy can be written as
\begin{align}\label{Action112CurvedPhB}
 E(\lambda)=\frac{I_2}{\lambda^3},
\end{align}
and we obtain, as in the standard case,
\be
\begin{split}\label{Stab_Ph_B}
&\frac{\partial E(\lambda)}{\partial \lambda}\bigg|_{\lambda=1}=0\rightarrow -3I_2= 0,\\
&\frac{\partial^2 E(\lambda)}{\partial \lambda^2}\bigg|_{\lambda=1}=12I_2=0\,,
\end{split}
\ee
which is again an inflection. 

In the case of Quintom fields, we have two interacting fields one canonical and the other non canonical. The action reads
\begin{equation}\label{eq:actionQuim}
\begin{split}
S=\frac{1}{2}\int d x^{4}\sqrt{-g}\Big[R&
-\nabla_a\varphi\nabla^a\varphi+ \\&
+\nabla_a\psi\nabla^a\psi- 2V(\varphi, \psi)
\Big]\;.
\end{split}
\end{equation}
Excluding backreaction,  the energy function associated to \eqref{eq:actionQuim} this action transforms under \eqref{QCBeta} as
\begin{align}\label{Action112CurvedQuimB}
 E(\lambda)=
&\left(\frac{I_1}{ \lambda}-\frac{I_4}{ \lambda}+\frac{I_2}{\lambda^3}\right),
\end{align}
where
\begin{equation}
\begin{split}
I_1&=\int e^{\Phi}\left[\varphi_{,q}^2+\sum_{i=2}^{3}\varphi_{,w_i}^2\right]\,  d\Omega \,,\\ 
I_2&=2\int e^{\Phi}V\left[\varphi(q)\right]\,  d\Omega\,,\\
I_4&=\int e^{\Phi}\left[\psi_{,q}^2+\sum_{i=2}^{3}\psi_{,w_i}^2\right]\,  d\Omega.
\end{split}
\end{equation}
This leads to 
\be
\begin{split}
&\frac{\partial E(\lambda)}{\partial \lambda}\bigg|_{\lambda=1}=0\rightarrow I_4= 3I_2+I_1,\\
&\frac{\partial^2 E(\lambda)}{\partial \lambda^2}\bigg|_{\lambda=1}=6I_2>0\,,
\end{split}
\ee
which, as in the case of the phantom field, can be stable if $I_2$ is positive. Again, using the trace of the gravitational field equations to include backreaction, we can write the action above on shell
\begin{equation}\label{eq:actionQuim_Shell}
\begin{split}
S=\int d x^{4}\sqrt{-g}V(\varphi, \psi)\;,
\end{split}
\end{equation}
which leads to instability. This was an expected result, which confirms the general conclusions we have drawn in Section \ref{secV}: a multi-field system becomes unstable if one of its components presents instability.

In the case of k-essence, the action is generalised as 
\begin{equation}\label{eq:actionK-ess}
\begin{split}
S=\frac{1}{2}\int d x^{4}\sqrt{-g}\Big[R+
P(\varphi,X)\Big]\;,
\end{split}
\end{equation}
where $X=\nabla_a\varphi\nabla^a\varphi$. Using the fact that for $\lambda=1$, ${ P}_{,\lambda}=2X{P}_{,X}$, under \eqref{QCBeta} and without backreaction one has
\be
\frac{\partial E(\lambda)}{\partial \lambda}\bigg|_{\lambda=1}=0 \rightarrow 2 X \partial_{X}P-3P=0\,,
\ee
which means
\be
P=P_0(\varphi)X^{3/2}. 
\ee
With this result one obtains 
\be
\begin{split}
&\frac{\partial^2 E(\lambda)}{\partial \lambda^2}\bigg|_{\lambda=1}=0.
\end{split}
\ee
In other words we always have instability.  

Considering backreaction, we have
\be
\frac{\partial E(\lambda)}{\partial \lambda}\bigg|_{\lambda=1}=0 \rightarrow  X \partial_{X}P-P=0,
\ee
which implies another inflection 
\be
\frac{\partial^2 E(\lambda)}{\partial \lambda^2}=0,
\ee
and therefore again instability.

On top of it intrinsic value, this result shows that to conditions of Derrick's theorem can be used to constrain modifications of general relativity in which undetermined functions are present. In the next section we will look at some interesting examples of such constraints.   
\section{Scalar tensor gravity}\label{NMC}
Using the results from the previous sections  we can proceed to the generalisation of Derrick's theorem to non minimal couplings. Let us consider, for example, the case of scalar tensor theories. This class of theories of gravity are characterised by 
\begin{equation}\label{eq:actionScTn}
\begin{split}
S=\frac{1}{2}\int d x^{4}\sqrt{-g}\Big[F(\varphi)R&
-\nabla_a\varphi\nabla^a\varphi 
- 2V(\varphi)
\Big]\;,
\end{split}
\end{equation}
whose variation gives 
the field equations
\be\label{FE_ST}
\begin{split}
FG_{ab}=&\nabla_{a}\varphi\nabla_{b}\varphi-g_{ab}\left[\frac{1}{2}\nabla_a\varphi\nabla^a\varphi + V(\varphi)\right]\\
&+ \nabla_{a}\nabla_{b}F -g_{ab}\dal F\,,
\end{split}
\ee
and the Klein-Gordon equation
\be\label{KleinST}
\dal\varphi+\frac{1}{2} R F_{,\varphi}-V_{,\varphi}=0,
\ee
were $F$ represents the non minimal coupling of the geometry (the Ricci scalar) with the field $\varphi$. 

Notice that, since the Ricci scalar naturally enters in \eqref{KleinST}, there is no need to add by hand backreaction. We will then treat the full case writing the action on shell.

Using the trace of the gravitational field equations and the Klein-Gordon equation, \eqref{eq:actionScTn} can be written as 
\begin{equation}\label{eq:actionScTn_Shell}
\begin{split}
S=\frac{1}{2}\int d x^{4}\sqrt{-g}\Big[&
K(\varphi)\nabla_a\varphi\nabla^a\varphi+ W(\varphi)\Big]\;,
\end{split}
\end{equation}
where
\begin{equation}
\begin{split}
K(\varphi)=&-\frac{3(F_{,\varphi}^2-2 F F_{,\varphi\varphi})}{2F+3F_{,\varphi}^2}\;,\\
W(\varphi)=&\frac{VF_{,\varphi}^2- FF_{,\varphi}V_{\varphi}-2 F V}{2F+3F_{,\varphi}^2}\;.
\end{split}
\end{equation}
A this point, defining the integrals
\begin{equation}\label{I1_I2_I3defSCTN}
\begin{split}
I_1&=\int e^{\Phi}K(\varphi)\left[\varphi_{,q}^2+\sum_{i=2}^{3}\varphi_{,w_i}^2\right]\,  d\Omega , \\ 
I_2&=2\int e^{\Phi}W\left(\varphi\right)\, d \Omega, 
\end{split}
\end{equation}
we can write
\begin{equation}\label{Energy112CurvedSCTN}
 E(\lambda)=
\left(\frac{I_1}{ \lambda}+\frac{I_2}{\lambda^3}\right),
\end{equation}
which leads to
\be
\begin{split}\label{Stab_SCTN_B}
&\frac{\partial E(\lambda)}{\partial \lambda}\bigg|_{\lambda=1}=0\rightarrow I_1+3I_2= 0,\\
&\frac{\partial^2 E(\lambda)}{\partial \lambda^2}\bigg|_{\lambda=1}=2I_1\,,
\end{split}
\ee
therefore stability with a non minimal coupling is possible if $I_1>0$ and  $I_2<0$. This results allows to give some limits on the functions $F$ and $V$. In particular
\be
-\frac{1}{6}\varphi^2<F\leq \frac{\beta ^2 }{4 \alpha }\varphi ^2 +\beta  \varphi+\alpha,
\ee 
and 
\be
V_0\varphi^4<V\leq V_0 \left(\varphi-\frac{2\alpha}{\beta}\right)^{2-\frac{4\alpha}{3\beta^2}},
\ee 
where  $V_0<0$, $\alpha>0$ and $\beta$ can be chosen freely.

Notice that \eqref{eq:actionScTn_Shell} can only be used if we exclude the coupling 
\be
F_C=-\frac{1}{6}\varphi^2.
\ee 
In order to obtain a result also in this case we can construct an action on shell using the Klein Gordon equation only. We obtain
\begin{equation}
\begin{split}
S=-\frac{1}{2}\int d x^{4}\sqrt{-g}\Big[&\varphi\dal\varphi
+\nabla_a\varphi\nabla^a\varphi+2 W_C(\varphi)\Big]\;,
\end{split}
\end{equation}
where
\be
W_C(\varphi)=V- \varphi V_{,\varphi}.
\ee
The above is in principle a non canonical action. However, it can be converted in a canonical one integrating by parts the higher order term. Indeed
\be
\varphi\dal\varphi= \nabla_a\left(\varphi\nabla^a\varphi\right)-\nabla_a\varphi\nabla^a\varphi.
\ee
Thus the action on shell can be written as
\begin{equation}\label{eq:actionScTn_Conf_Shell}
\begin{split}
S=-\frac{1}{2}\int d x^{4}\sqrt{-g}\Big[&2\nabla_a\varphi\nabla^a\varphi+ W_C(\varphi)\Big]\;.
\end{split}
\end{equation} 
In this way we can employ the usual procedure to explore this case. We have
\begin{equation}
 E(\lambda)=
\left(2\frac{I_1}{ \lambda}+\frac{I_2}{\lambda^3}\right),
\end{equation}
with
\begin{equation}\label{I1_I2_I3defSCTN_Conf_Shell}
\begin{split}
I_1&=\int e^{\Phi}\left[\varphi_{,q}^2+\sum_{i=2}^{3}\varphi_{,w_i}^2\right]\,  d\Omega , \\ 
I_2&=2\int e^{\Phi}W_C\left(\varphi\right)\, d \Omega, 
\end{split}
\end{equation}
which leads to
\be
\begin{split}\label{Stab_SCTN_B_Conf_Shell}
&\frac{\partial E(\lambda)}{\partial \lambda}\bigg|_{\lambda=1}=0\rightarrow 2I_1+3I_2= 0,\\
&\frac{\partial^2 E(\lambda)}{\partial \lambda^2}\bigg|_{\lambda=1}=-4I_1<0\,,
\end{split}
\ee
and implies instability. All in all therefore only very specific combination of coupling and potential can lead to stable scalar field configurations.

It is widely believed today that the most general model with a single additional scalar degree of freedom  and second order field equations is given by the so called Horndeski theory \cite{Horndeski:1974wa}. This theory has been proved to be equivalent to the  curved spacetime generalisation of a scalar field theory with Galilean shift symmetry in flat spacetime  \cite{Deffayet:2009mn,Kobayashi:2011nu}. We will now apply the equilibrium  and stability conditions we have derived above to this class of theories.

It is useful for our purposes to write their action in the form
\begin{equation}
S=\sum_{i=1}^{4}\int d^{4}x\sqrt{-g}{\cal L}_i\,,
\end{equation}
where
\begin{align}
{\cal L}_1=&G_1\,,\label{g2}\\
{\cal L}_2=&-G_{2}\dal\varphi\,,\\
{\cal L}_3=&G_{3}R+G_{3,X}\left[(\dal\varphi)^2-\varphi_{; mn}\varphi^{; mn}\right]\,,\\
{\cal L}_4=&G_{4}\,\mathds{G}_{m n}\varphi^{; mn}-\frac{{G}_{4,X}}{6}\left[(\dal\varphi)^3 \right.\\
&\left.+2\varphi_{; mn}\varphi^{;m}{}_{;a} \varphi^{; an}-3\varphi_{; mn}\varphi^{; mn}\dal\varphi\right]\,.
\end{align}
Here $\mathds{G}_{m n}$ is the Einstein tensor, $G_{i}$ are functions of $X$ and $\varphi$. Under \eqref{QCBeta} we have, together with \eqref{R_transf},
\be
\begin{split}
&\dal\varphi\Rightarrow \lambda^2 \dal\varphi + ...\\
&\varphi_{; mn}\Rightarrow\varphi_{; mn}+ ...\\
&\mathds{G}_{m n}\varphi^{; mn}\Rightarrow \lambda^6 \mathds{G}_{m n}\varphi^{; mn} + ...
\end{split}
\ee
where, as before, the dots represent additional terms which contain derivatives of $\lambda$ and therefore are irrelevant in our case. 

Before proceeding we should point out  that this theory has a non canonical Lagrangian and therefore the energy function is not the standard one. However, the Ostrogradski approach \cite{Ostrogradsky:1850fid} can be used to show that also in this case one can define a function that has the characteristics of the energy of the system (i.e. it is conserved and generates a time evolution).

The transformed action, in the static case and assuming for brevity $\lambda=const.$, can be written as
\begin{equation}
\begin{split}
S(\lambda)&=\int d^{4}x\frac{\sqrt{-g}}{\lambda^3}\left[{\cal L}_1+\lambda^2{\cal L}_2+\lambda^4{\cal L}_3+\lambda^6{\cal L}_4\right]\,,\\
\end{split}
\end{equation}
where ${\cal L}_i={\cal L}_i(\lambda)$. 

The Pohozaev identity for $\lambda=1$ reads
\be\label{HornPhzv}
\begin{split}
&2X{\cal L}_{1,X}-3{\cal L}_1+2X{\cal L}_{2,X}-{\cal L}_2\\
&+2X{\cal L}_{3,X}+{\cal L}_3+2X{\cal L}_{4,X}+3{\cal L}_4=0\,.
\end{split}
\ee
The \eqref{HornPhzv} contains derivatives of the scalar field and therefore, without further assumptions, it cannot be used to obtain general constraints of the functions $G_i$ as we have done in the previous section. 

When such assumptions are provided one can find a number different combinations of these functions which can lead to stability. The trivial ones are the ones corresponding to  general relativity ($G_{1}=-\frac{1}{2}X-V$,$G_{2}=0$, $G_{3,X}=0$ and $G_{4}=0$) and scalar tensor gravity ($G_{1}=-\frac{1}{2}C_1(\varphi)X-V$,$G_{2}=0$, $G_{3}=C_3(\varphi)$ and $G_{4}=0$).
A more general analytical condition can be obtained, for example, assuming only $G_{4}=0$. In this case stability is possible for
\be
\begin{split}
G_2&=C_{2,1}(\varphi)X^{1/2}+C_{2,2}(\varphi)X,\\
G_3&=C_{3,1}(\varphi)X^{3/2}+C_{3,2}(\varphi),\\
\end{split}
\ee
and $G_1$ within the functions
\be
\begin{split}
G^*_{1}=&C_{1,1}(\varphi)X^{5/2}+C_{1,2}(\varphi)X^{3/2}+C_{1,3}(\varphi)X\\
G^*_2=&C_{1,5}X^{3/2}+C_{1,6}(\varphi)X+C_{1,4}(\varphi)\ln X,\\
&+C_{1,7}(\varphi).\\
\end{split}
\ee
Note that this is only a particular solution  for the $G_i$. Obtaining a general solution would require the resolution of a non linear third order differential equation for $G_3$ which cannot be achieved in general. 

As said, there is a number of other combinations of forms of the function $G_i$ which can lead to stability. The fact that such a big number of different conditions is possible is to be ascribed to the high level of generality of this class of theories. Members of Horndeski group of theories can have a wildly different physical behaviour and this reflects also in the stability of localised solution of their scalar degree of freedom.

\section{Conclusion}\label{conclusions}

In this paper we have presented a complete proof of Derrick's theorem in curved spacetime. 
The proof follows the same strategy of the original paper by Derrick, but has the advantage to be completely covariant and to not require any assumption on the potential for the scalar field nor  on the underlying geometry (other than the absence of vorticity).  This is made possible by combining the 1+1+2 covariant formalism and a technique firstly developed by Darboux which allows to write down Lagrangians for dissipative systems.

For scalar fields in a fixed background, i.e. non backreacting, we have been able to prove that  no stable localised solution of the Klein-Gordon equation is possible.  In addition, we found  that the coupling of the scalar field with other types of matter can only change the conditions necessary to achieve equilibrium. 

The results we have obtained can be understood recognising that a scalar field can be represented macroscopically as an effective fluid with negative pressure (tension). Such fluids will tend naturally to collapse in flat spacetime. In the non flat case, if the energy of the scalar field is not enough to appreciably influence the curvature of spacetime (i.e. without backreaction) a localised solution will be again unstable. When couplings are considered, the  interaction influences the tension of the scalar field, changing its magnitude.

In the case of real scalar field stars, as the metric of these compact objects is determined by its matter distribution, backreaction cannot be neglected. Our generalisation of Derrick's theorem, therefore, implies that in curved spacetimes stable relativistic stars made of real scalar fields cannot exist, even when considering couplings. Indeed the introduction of gravity ``mitigates'' the instability in the sense that the maximum of the energy of the scalar field solution is turned in an inflection point.

In terms of the fluid interpretation this is also clear: if the scalar field is able to influence the spacetime in which it is embedded, its tension will induce a repulsion, much in the same way as dark energy. As a result, the configuration is less unstable and, in principle, with enough tension stability could be possible. Our result suggests, however, that no sufficient level of tension can be achieved to support a localised solution. 

It should be remarked, however, that full stability is not a necessary conditions for physical validity. In order to be able to consider scalar field stars as possible astrophysical objects one can require the weaker conditions of ``long term'' stability (e.g. longer/comparable to the age of the Universe). It is not possible with sole the tools of Derrick's proof to evaluate this aspect of the stability of scalar field stars. Its study will be the topic of future works.

Differently from full stability, the equilibrium condition still constitutes a strong physical constraint. Its  application in the context of non canonical scalar field and some classes of modified gravity reveals that the extended Pohozaev identity can be used to select potentials and classes of theories which present an equilibrium. This is particularly relevant in the case of modified gravity, as for these theories all compact stars are also real scalar field stars in which the scalar field coincides with the gravitational scalar degrees of freedom. We found some very stringent criteria for scalar tensor gravity and Horndeski-type theories.  Also in this case, coupling with other fields as well as standard matter might be able to modify these constraints.

We conclude pointing out that our analysis was performed in a purely classical setting: we have neglected completely the quantum nature of the scalar field and the corresponding modification to its action. These corrections, which are necessary to build more realistic models of scalar field stars, could lead to modifications of the equilibrium conditions we have derived and even to stability. Future studies will be dedicated to the analysis of these cases.

\appendix

\section{Connecting $\sqrt{-g}$ and $e^\Phi$}\label{app}
Let us start with the covariant divergence of the vector $u_a$. It is well known that
\begin{equation}
\nabla_a u^a= \frac{1}{\sqrt{-g}}\partial_a\left(\sqrt{-g} u^a\right).
\end{equation}
On the other hand, by definition, $\nabla_a u^a=\Theta$  i.e. the left hand side is the expansion. In this way one can write
\begin{equation}
u^a\partial_a \ln |g|= 2(\Theta-\partial_a u^a).
\end{equation}
The same procedure can be applied to the quantity $\nabla_a e^a$ to obtain
\begin{equation}
 e^a\partial_a \ln |g|= 2(\A+\phi-\partial_a e^a),
\end{equation}
Instead, for $N_{ca}\nabla_{b}N^{ab}$, since
\begin{equation}
\nabla_b N^{ab}= \frac{1}{\sqrt{-g}}\partial_b\left(\sqrt{-g} N^{ab}\right)+\Gamma_{bc}^a N^{bc},
\end{equation}
we have
\begin{equation}
N_c{}^a \partial_a \ln |g|=  2(\A_c+a_c)+N_c{}^b \partial_b \ln{N},
\end{equation}
where we call $N$ the determinant of the non zero minor of $N_{ab}$. In this way we have
\begin{equation}\label{eq_g}
\begin{split}
\partial_a \ln |g| = &2\Big\{(\partial_c u^c-\Theta)u_a+(\A+\phi-\partial_c e^c)e_a\\ & +\left(\A_c+a_c\right)N^c_a\Big\}+N_a{}^b \partial_b\ln{N}.
\end{split}
\end{equation} 
Let us now define
\begin{equation}
\begin{split}
V_a = &-\Theta u_a+(\A+\phi)e_a+\left(\A_c+a_c\right)N^c_a,\\ 
W_a = &-\left(\partial_b u^b\right) u_a+\left(\partial_b e^b\right) e_a+N_a{}^b \partial_b \ln{\sqrt{N}},
\end{split}
\end{equation} 
so that 
\begin{equation}\label{der_g}
\partial_a \ln{|g|} =2(V_a- W_a).
\end{equation}
The partial differential equation \eqref{eq_g} is equivalent to the one we have encountered in the main text to determine $\Phi$ (see \eqref{SolPhiSS} for $\lambda=1$). 

In the case of a static and spherically symmetric spacetime and choosing $e_a$ normalised to one we have $\Theta=0$, $\A_c=0$, $a_c=0$ and $\partial_a u^a= 0$. Inserting this result in \eqref{der_g} yields and integrating out the total divergences in $W_a$ we obtain
\begin{equation}\label{SolPhiMet}
\sqrt{|g|}= g_0\exp\left[\int \left(\A+\phi\right)d q\right]\,,
\end{equation}
i.e., modulus an irrelevant constant, the \eqref{Phi_Beta}. This suggests that we can write in a $(-,+,+,+)$ signature
\begin{equation}
e^\Phi =\sqrt{-g}.
\end{equation}
In more general spacetimes \eqref{der_g} is a system of partial differential equations which one must solve in order to find the expression of the volume form. As the case of equation \eqref{EqDarboux112Curv}, we can use the method of characteristics to obtain some solutions, but we need an accurate description of the boundary conditions to determine a solution. However, as discussed for $\Phi$, the exact form of the metric tensor is irrelevant for our discussion, the only important thing is the transformation of $\sqrt{-g}$. 

Let us then look at the transformation of $g$. It is easy to see that under a conformal transformation $g_{ab}(\lambda)=\lambda g_{ab}$ one has
\begin{equation}
\begin{split}
V_a(\lambda) = &V_a\,,\\ 
W_a(\lambda) = & W_a + \partial_a \ln\lambda,
\end{split}
\end{equation}
and this implies 
\begin{equation}
\Phi (\lambda)
=\Phi- 4 \ln \lambda\,,
\end{equation}
\begin{equation}
g (\lambda)
=\frac{g}{\lambda^4}\,,
\end{equation}
which is consistent with the known conformal transformation of a tensor density.

Under \eqref{QCBeta} one has, again,
\begin{equation}
\begin{split}
V_a(\lambda) = &V_a\,,\\ 
W_a(\lambda) = & W_a + \partial_a \ln\lambda\,.
\end{split}
\end{equation}
However, the above results imply 
\begin{equation}
\begin{split}
\Phi (\lambda)
&=\Phi- 3 \ln \lambda,\\
g (\lambda)
&=\frac{g}{\lambda^3},
\end{split}
\end{equation}
i.e. the same transformation for $\Phi$ obtained in \eqref{SolPhiSS}.
\medskip

\begin{acknowledgments}
SC~was supported by  the Funda\c{c}\~{a}o para a Ci\^{e}ncia e Tecnologia through project IF/00250/2013 and partly funded through H2020 ERC Consolidator Grant ``Matter and strong-field gravity: New frontiers in Einstein's theory'', grant agreement no. MaGRaTh-64659.
JLR acknowledges financial support of FCT-IDPASC through grant no. PD/BD/114072/2015. 
\end{acknowledgments}


\end{document}